\documentclass[aps,prc,amsmath,amssymb,amsfonts,reprint,showpacs,showkeys]{revtex4-1}
\usepackage{graphicx}
\begin{document}
\title{Modification of Eckart theory of relativistic dissipative fluid\\
by introducing extended matching conditions}

\author{Takeshi Osada}
\email[]{osada@ph.ns.tcu.ac.jp}

\affiliation{
Department of Physics, School of Liberal Arts, Tokyo City University,
Tamazutsumi 1-28-1, Setagaya-ku, Tokyo 158-8557, Japan}

\date{\today}
\begin{abstract}
We deal with a novel approach to formulation of the relativistic dissipative hydrodynamics 
by extending the so-called matching conditions widely used in the literature. 
The form of the non-equilibrium entropy current 
can be determined by requiring  thermodynamical stability of the entropy current 
under extended matching conditions. We derive equations of motion for 
the relativistic dissipative fluid based on the Eckart theory and show that linearized equations obtained from 
them are stable against small perturbations.  
It is also shown that the required fluid stability conditions are related to the causality of the model.  
\end{abstract}

\maketitle
%
\section{Introduction}

Israel and Stewart (IS) provide phenomenological framework for 
a relativistic dissipative fluid \cite{Israel1976310,Israel1979341}. 
In their model, a possible general form for the non-equilibrium entropy current is 
described by the dissipative part of the energy-momentum tensor and by the  
particle current up to the second order in the deviation from the equilibrium state. 
After the stability and causality of the IS theory have been shown 
by Hiscock and Lindblom \cite{Hiscock1983466,PhysRevD.31.725,PhysRevD.35.3723},
the causal dissipative hydrodynamical model was adapted to study the dynamics of 
hot matter produced in ultra-relativistic heavy-ion 
collisions by Muronga \cite{PhysRevLett.88.062302,PhysRevC.69.034903}. 
In recent years, the IS model plays important role 
in the analysis of the experimental data obtained by RHIC and LHC (see for example \cite{PhysRevC.84.027902}).\\

In parallel to application of IS theory, 
investigations of the basis of relativistic dissipative hydrodynamic theory 
was vigorously continued \cite{PhysRevC.75.034909,
PhysRevD.81.114039,PhysRevLett.105.162501,0954-3899-35-11-115102,0954-3899-36-3-035103,
Tsumura2007134,Tsumura2010255, Van:2011yn, springerlink:10.1140/epjst/e2008-00602-6}. 
The reason is that the IS theory in its present form is too general and complex from the point of view 
of the quantum chromodynamics (QCD), which is belived to be the fundamental theory for strongly 
interacting system \cite{PhysRevC.75.034909}. 
Another reason is that theory of the relativistic dissipative hydrodynamics is not yet fully understood because, 
for example, the equation of motion of the fluid it uses depends on the choice of the Lorentz frame 
(or on the definition of the hydrodynamical flow \cite{Tsumura2007134}).  
Since the dissipative part of the energy-momentum tensor $\delta T^{\mu\nu}$ 
and the particle current $\delta N^{\mu}$ cannot be 
determined uniquely by the second law of thermodynamics, one usually 
introduces some constraints to fix them, they are known as matching conditions.\\ 

The matching (or fitting) conditions are also introduced because of necessity of 
{\it matching} the energy density and  baryonic charge density $(\varepsilon, n)$ 
in a non-equilibrium state to the corresponding equilibrium densities:  
($\varepsilon_{\rm eq}$, $n_{\rm eq}$), 
$\varepsilon = \varepsilon_{\rm eq},~n=n_{\rm eq}$, or equivalently,  
\begin{eqnarray}
  u_{\mu} u_{\nu }\delta T^{\mu\nu}  =0, \quad \delta u_{\mu} N^{\mu} =0. 
\label{eq:standard_matching_condition}
\end{eqnarray} 
These matching conditions allows to determine the thermodynamical 
pressure $P_{\rm eq}(\varepsilon_{\rm eq},n_{\rm eq})$ (defined as work done in isentropic expansion) 
via equation of state for the equilibrium state. 
Here the $P_{\rm eq}$ should be distinguished from the bulk viscous contribution,  
$\Pi\equiv -\frac{1}{3}\Delta_{\mu\nu}\delta T^{\mu\nu}$, present 
in the energy-momentum tensor \cite{Israel1979341}.  
The matching conditions are also needed because they are necessary for the thermodynamical stability 
of the entropy current (see Appendix A in Ref.\cite{PhysRevC.80.054906}). 
However, the matching condition given by eq.(\ref{eq:standard_matching_condition}) are not unique. 
So far, except of some recent works \cite{Tsumura2007134,Tsumura2010255}, 
they were not investigated in detail. 

A state of  relativistic dissipative fluid is described by energy-momentum tensor $T^{\mu\nu}(x)$ 
and by the baryon number current $N^{\mu}(x)$, which obey the conservation laws 
\footnote{The notation ${X}^{\lambda}_{;\mu}$ denotes the covariant derivative of the 
vector $X^{\lambda}$ compatible with the space-time metric $g^{\mu\nu}$. 
For a scalar $X$,  or in the Minkowski space-time ($g^{\mu\nu}=\mbox{diag}(1,-1,-1,-1)$), 
the covariant derivative reduces to the usual 
derivative, $\partial_{\mu} X$. We hereafter denote $X_{,\mu}$ such 
derivative distinctly.}
\begin{subequations}
\begin{eqnarray}
  && T^{\mu\nu}_{;\mu} = 0, \label{eq:FluidEquation} \\ 
  && N^{\mu}_{;\mu} =0,\label{eq:ChargeConservation} 
\end{eqnarray}
and the second law of the thermodynamics 
\begin{eqnarray} 
  && S^{\mu}_{;\mu}\ge 0. \label{eq:2nd_law}
\end{eqnarray}
\end{subequations}
Because of the uncertainty in definition of the flow velocity $u^{\mu}(x)$ for a non-equilibrium fluid, 
one needs, unlike in the case of perfect fluid, to fix the frame for the fluid considered. 
Throughout this paper we employ the so-called Eckart frame \cite{PhysRev.58.919}. 
It means that the hydrodynamical flow velocity $u^{\mu}$ (with normalization $u^{\mu}u_{\mu}=1$)  
is defined by using baryon charge current     
\begin{eqnarray}
  u^{\mu}(x) \equiv \frac{N^{\mu}}{\sqrt{N^{\nu} N_{\nu}}}.
\label{eq:FluidVelocityEckart}
\end{eqnarray}
In this frame one has always $\Delta^{\mu}_{\lambda} N^{\lambda} =0$ with 
$\Delta^{\mu}_{\nu} \equiv g^{\mu}_{\nu} -u^{\mu}u_{\nu}$ 
being the projection operator orthogonal to the four vector $u^{\mu}$.
Although the alternative Landau-frame \cite{Landau} (in which the nonzero baryon number 
current exists $\Delta^{\mu}_{\lambda}N^{\lambda}\ne 0$ but the energy flow disappears, 
$W^{\mu}\equiv u_{\nu}\Delta^{\mu}_{\nu} T^{\nu\lambda} =0$) is 
more relevant in the context of ultra-relativistic heavy-ion collisions, 
we use the Eckart frame because of our previous work \cite{PhysRevC.81.024907}, in which 
we have argued that for the relativistic fluid in the presence of the long-range correlations 
$u_{\mu}u_{\nu}\delta T^{\mu\nu} \ne 0$. 

In the present work we reformulate the relativistic dissipative hydrodynamics 
based on Eckart prescription \cite{PhysRev.58.919} using following 
{\it extended matching conditions},  
\begin{eqnarray}
  u_{\mu} u_{\nu} \delta T^{\mu\nu}  ={\cal O}(1), \quad u_{\mu} \delta N^{\mu}  ={\cal O}(2), \label{eq:generalized_matching}
\end{eqnarray}
instead of those given by eq.(\ref{eq:standard_matching_condition}). 
Here, ${\cal O}(1)$ denotes the first order infinitesimal quantities, i.e., the difference of fluid dynamical 
quantities between the equilibrium and the dissipative state 
(for example, bulk pressure $\Pi\equiv P-P_{\rm eq}$ is of the ${\cal O}(1)$ order).   
On the other hand,  $u_{\mu} \delta N^{\mu}$ is assumed to be of the order ${\cal O}(2)$ 
because of introduction of infinitesimal ${\cal O}(2)$ to the condition $u_{\mu}\delta N^{\mu}=0$; 
it is sufficient to derive the equations of motion for stable fluid as discussed in \ref{sec.III}. 

Hence, the main purpose of this article is to investigate the system  described by equations 
eq.(\ref{eq:FluidEquation}) and eq.(\ref{eq:ChargeConservation}), 
in the frame defined by eq.(\ref{eq:FluidVelocityEckart}), 
while observing eq.(\ref{eq:2nd_law}), 
with constraints provided by (\ref{eq:generalized_matching}) instead of eq.(\ref{eq:standard_matching_condition}). \\

This article is organized as follows. In Sec.\ref{sec.II}, we briefly
review the relativistic fluid dynamics with standard matching conditions. 
Then extended matching conditions are 
introduced into the hydrodynamical model using 
general form of an irreversible entropy current. 
In Sec.\ref{sec.III}, we check the stability and causality of 
the fluid obtained from our model.   
We close with Sec. \ref{sec.IV} containing 
the summary and some further discussion. 
Some technical details are presented in Appendix A and B. 

\section{Relativistic dissipative hydrodynamics with extended matching conditions}\label{sec.II}
\subsection{Eckart theory with standard matching conditions} 
The entropy current can be written in the following simple form 
\begin{eqnarray}
S^{\mu}(x)
&\equiv& 
 -\alpha \phi^{\mu} +  \beta_{\lambda}\Phi^{\lambda\mu}, 
\end{eqnarray} 
where $\phi^{\mu}$ and $\Phi^{\mu\nu}$ are, respectively, vector and  
2-rank symmetric tensor, $\alpha\equiv \mu_b/T$ and $\beta^{\mu}=\beta u^{\mu}$ with 
$\beta\equiv 1/T$, $T$ and $\mu_b$ being, respectively, the temperature and baryon chemical potential.  
In a local equilibrium case it is then given by 
\begin{subequations}\label{eq:equilibrium_current} 
\begin{eqnarray}
S^{\mu}_{\rm eq} &\equiv& -\alpha\phi^{\mu}_0 +\beta_{\lambda}\Phi^{\lambda\mu}_0, \label{eq:equilibrium_entropy_current}\\ 
    \phi^{\mu}_0 &=&   N^{\mu}_{\rm eq} ,\quad 
     \Phi^{\lambda\nu}_0 = T_{\rm eq}^{\lambda\nu} 
     - \frac{g^{\lambda\nu}}{3}\Delta_{\alpha\beta}T^{\alpha\beta}_{\rm eq}, 
     \label{eq:phi0PHI0}
\end{eqnarray}
\end{subequations} 
where $T^{\mu\nu}_{\rm eq}$ and $N^{\mu}_{\rm eq}$ are equilibrium energy-momentum tensor and baryonic charge current: 
\begin{subequations}
\begin{eqnarray}
  T^{\mu\nu}_{\rm eq} &=& \varepsilon_{\rm eq} u^{\mu}u^{\nu} -P_{\rm eq}\Delta^{\mu\nu}, \\
  N^{\mu}_{\rm eq} &=& n_{\rm eq} u^{\mu}. 
\end{eqnarray}
\end{subequations} 
In the equilibrium case, 
the energy-momentum conservations $T^{\mu\nu}_{{\rm eq};\mu}=0$,  
and the baryon number conservation $N^{\mu}_{{\rm eq};\mu}=0$, 
together with thermodynamic relations, result in the locally conserved entropy current: 
\begin{eqnarray}
  S^{\mu}_{{\rm eq};\mu} &=& -\alpha_{,\mu}N^{\mu}_{\rm eq} 
  +\beta_{\lambda;\mu} T^{\lambda\mu}_{\rm eq} +[\beta^{\mu}P_{\rm eq}]_{;\mu} 
=0. 
\end{eqnarray}
To this current we introduce now dissipative corrections by adding corresponding dissipative corrections 
$\delta T^{\mu\nu}$ and $\delta N^{\mu}$ to the energy-momentum tensor and to baryon number current   
appearing in $\phi_0^{\mu}$ and $\Phi_0^{\mu\nu}$ in eq.(\ref{eq:equilibrium_current}). 
In this way one extends 
the expression of equilibrium entropy current towards the off-equilibrium entropy current, 
\begin{subequations}
\begin{eqnarray} \label{eq:general_entropy}
&& S^{\mu} \equiv -\alpha[\phi^{\mu}_0+\delta \phi^{\mu}] 
    +\beta_{\lambda}[\Phi^{\lambda\mu}_0+\delta\Phi^{\lambda\mu}], \label{eq:general_entropy1}\\ 
&& \delta \phi^{\mu} =\delta N^{\mu}, \quad  
     \delta \Phi^{\lambda\nu} = \delta  T^{\lambda\nu}  
     +\chi ~\frac{g^{\lambda\nu} }{3}\Delta_{\alpha\beta} \delta T^{\alpha\beta}. \quad
     \label{eq:general_entropy2}
\end{eqnarray}
\end{subequations} 
Notice that, 
because of the thermodynamical stability, 
term proportional to $\Delta_{\alpha\beta} \delta T^{\alpha\beta}$, 
which appears due to the natural extension of eq.(\ref{eq:phi0PHI0}),  
is usually dropped. 
However, since we shall generalize these conditions,  
we retain this term and introduce a phenomenological 
parameter $\chi$ which determines its contribution to the non-equilibrium entropy current. 
Note that in the Eckart frame $\delta N^{\mu}\equiv 0$. 
This is because $\Delta_{\lambda}^{\mu}\delta N^{\mu}=0$ 
(by definition of the Eckart frame)  
and because of one of the matching conditions, $u_{\mu} \delta N^{\mu} =0$.    
From another matching conditions, $u_{\mu}u_{\nu} \delta T^{\mu\nu} $=0, 
it follows that 
one can also drop term proportional to $u^{\mu}u^{\nu}$. 
As result one gets  
\begin{eqnarray}
 \delta T^{\mu\nu} = -\Pi\Delta^{\mu\nu} + W^{\mu}u^{\nu}+W^{\nu}u^{\mu} +\pi^{\mu\nu}, \label{eq:Tmunu_standard}
\end{eqnarray}
where $\Pi$ is bulk pressure, $W^{\mu}$ is energy flow and $\pi^{\mu\nu}$ is shear tensor. 
Using now this expression for $\delta T^{\mu\nu}$ in eq.(\ref{eq:general_entropy2}) 
we obtain following possible expression for 
the non-equilibrium entropy current  
\begin{eqnarray}
 S^{\mu} &=& -\alpha \phi^{\mu} +\beta_{\lambda} \Phi^{\lambda\mu} 
 \nonumber \\ &=&  
 S^{\mu}_{\rm eq} + \frac{W^{\mu}}{T} -\chi\frac{\Pi}{T}u^{\mu} . \quad 
 \label{eq:Smu_1st}
\end{eqnarray}
In the usual formulation, i.e., for standard matching conditions, 
term proportional to $\Pi u^{\mu}$ is dropped 
to prevent the occurrence of instability in the entropy current 
(in our case it would mean assuming $\chi=0$).  
In this case one gets the known first order theory form of the entropy current \cite{PhysRev.58.919}: 
\begin{eqnarray}
 S^{\mu} \mapsto S^{\mu}_{\rm E}  = S^{\mu}_{\rm eq} + \frac{W^{\mu}}{T}. \label{eq:CurrentEckart}
\end{eqnarray}
Correspondingly, the entropy production in the Eckart's 1st order theory is given by 
\begin{eqnarray}
  S^{\mu}_{{\rm Ec};\mu} 
  &=& \frac{u_{\nu}}{T}T^{\mu\nu}_{{\rm eq};\mu} + [\frac{u_{\nu}}{T}\delta T^{\mu\nu}]_{;\mu}  
  = [\frac{u_{\nu}}{T}]_{;\mu} \delta T^{\mu\nu} .\label{eq:Eckart's_production}
\end{eqnarray} 
However, for more general form of matching conditions proposed here this term should be included with 
parameter $\chi$ used to restrict the form of the entropy current in the non-equilibrium state. 

\subsection{Extended matching conditions}
In what follows, we propose and discuss in detail the {\it extended matching conditions},   
\begin{subequations}
\begin{eqnarray}
u_{\mu} u_{\lambda} \delta T^{\lambda\mu}   &=& \Lambda, 
\label{eq:ExMatching_Tmunu} \\
u_{\sigma}\delta N^{\sigma} &=& \delta n , 
\label{eq:ExMatching_Nmu} 
\end{eqnarray}
\end{subequations}
to be used instead of the standard matching conditions given by 
eq.(\ref{eq:standard_matching_condition}) and widely used in the literature.  
In this case the dissipative part of energy-momentum tensor is give by 
\begin{eqnarray}
 \delta T^{\mu\nu} = 
 \Lambda u^{\mu}u^{\nu}-\Pi\Delta^{\mu\nu} + W^{\mu}u^{\nu} + W^{\nu}u^{\mu} + \pi^{\mu\nu}\quad 
\end{eqnarray}
instead of eq.(\ref{eq:Tmunu_standard}). 
Correspondingly, the entropy current eq.(\ref{eq:general_entropy}) is modified and takes form 
\begin{eqnarray}
S^{\mu} &=& -\alpha \phi^{\mu} +\beta_{\lambda} \Phi^{\lambda\mu} \nonumber \\
&=&  S^{\mu}_{\rm eq} + \frac{W^{\mu}}{T} -\alpha \delta n u^{\mu} +\Lambda \beta^{\mu}  -\chi\beta\Pi u^{\mu}. 
\label{eq:entropy_current_our_model} 
\end{eqnarray}
If we now require that 
\begin{eqnarray*}
 -\alpha \delta n u^{\mu} +\Lambda \beta^{\mu}  -\chi\beta\Pi u^{\mu} =0,  
\end{eqnarray*}
or equivalently, that 
\begin{eqnarray}
 -\mu_b\delta n  +\Lambda  - \chi \Pi =0, \label{eq:StabilityCondition}
\end{eqnarray}
the entropy current reduces to its usual Eckart's form, $S^{\mu}\mapsto S^{\mu}_{E}$.  
Eq.(\ref{eq:StabilityCondition}) defines therefore new thermodynamical stability condition 
which replaces previous requirement that in eq.(\ref{eq:Smu_1st}) $\chi=0$; 
this new condition does not affect the local entropy density. 
On the other hand, extended matching condition, eq.(\ref{eq:ExMatching_Tmunu}), results in  
\begin{eqnarray*}
 W^{\mu}/T \ne \beta_{\lambda}\delta T^{\lambda\mu},
\end{eqnarray*}  
therefore now one gets 
\begin{eqnarray}
  S^{\mu} &=& S^{\mu}_{\rm eq}+\frac{W^{\mu}}{T}
  ~=S^{\mu}_{\rm eq}+\beta_{\lambda}\delta T^{\lambda\mu} -\Lambda \beta^{\mu}.
  \label{eq:entropy_current} 
\end{eqnarray} 
The entropy production is thus in this case given by 
\begin{eqnarray}
  S^{\mu}_{;\mu}  
  &=&  -\alpha_{,\mu}  \delta N^{\mu}  + \beta_{\lambda;\mu}\delta T^{\lambda\mu} 
          - [ \chi \Pi\beta^{\mu}]_{;\mu} \nonumber \\ 
  &=& S^{\mu}_{{\rm E};\mu} 
  +[ \alpha \frac{d \delta n}{d\tau}- \beta\frac{d\Lambda}{d\tau} ] 
  +[ \alpha \delta n -\beta \Lambda ] \theta,
  \label{eq:entropy_production}
\end{eqnarray}
where $\theta\equiv u^{\mu}_{;\mu}$ denoting the local expansion rate. 
The last two terms represent corrections to the usual Eckart's formula, 
eq.(\ref{eq:Eckart's_production}).

We assume now that $\Lambda$ and $\delta n$ can be 
expressed by the respective scalar quantities: $\Pi$ and $W^{\mu}W_{\mu}$ and 
$\pi^{\mu\nu}\pi_{\mu\nu}$. Furthermore, we assume that 
they are given by ${\cal O}(1)$ and ${\cal O}(2)$ infinitesimal quantities. 
In the most simple way they can be then written as:  
\begin{subequations}\label{eq:extended_matching}
\begin{eqnarray}
   && \Lambda =\kappa \Pi, \label{eq:extended_matching_1}\\ 
   && \mu_b \delta n = -\xi \Pi^2 + \xi' W^{\mu}W_{\mu}  -\xi'' \pi^{\mu\nu}\pi_{\mu\nu}, \label{eq:extended_matching_2}
\end{eqnarray} 
\end{subequations}
where $\kappa, \xi, \xi'$, and $\xi''$ are positive constants.
Note that $\xi, \xi'$ and $\xi''$ have dimension [GeV]$^{-4}$ whereas 
$\kappa$ has is dimensionless.  
It is interesting that  $\Lambda={\cal O}(1)$ ($\kappa=3$) is obtained 
in Ref.~\cite{Tsumura2010255} and also obtained independently in Ref.~\cite{PhysRevC.81.024907} 
(see also Ref.~\cite{PhysRevC.77.044903}). 
$\delta n={\cal O}(2)$ is needed to stabilize the so-called transverse mode propagation of the dissipative fluid 
(discussed below in Sec.\ref{sec:transverse_mode}). 

A comment concerning the matching conditions may be in order here. 
The IS theory introduces contributions of the second order, ${\cal O}(2)$, 
$Q^{\mu} ( \delta T^{\mu\nu}, \delta N^{\mu} )$   
to the entropy current (see eq.(2.17) in the ref.\cite{Israel1979341}). 
In the present model, the second order contributions to the entropy current are brought in by the 
matching condition eq.(\ref{eq:extended_matching_2}), $u_{\mu} \delta N^{\mu}\ne 0$ 
(together with contributions of the first order, brought by $u_{\mu}u_{\nu}\delta T^{\mu\nu}\ne 0$,  
eq.(\ref{eq:extended_matching_1}). 
Notice, however, that, because of the stability condition eq.(\ref{eq:StabilityCondition}), 
one has $Q^{\mu}\equiv 0$, i.e., terms introduced by extended matching conditions do not contribute 
to the total entropy current. 
In this way our model is, in fact, a first order model and because of this differs from the IS approach. \\ 

Notwithstanding the fact that out matching conditions do not give direct contribution to the entropy current, 
the resulting equation of motion of fluid changes its form accordingly.  
The thermodynamical stability condition, eq.(\ref{eq:StabilityCondition}), now reads
\begin{eqnarray}
  \chi=  \kappa +\xi\Pi   
          -\xi' \frac{W^{\mu}W_{\mu}}{\Pi} +\xi'' \frac{\pi^{\mu\nu}\pi_{\mu\nu}}{\Pi} . 
\end{eqnarray}
Because the entropy production eq.(\ref{eq:entropy_production}) can be rewritten as
\begin{eqnarray}
  T S^{\mu}_{;\mu} 
  &=& TS^{\mu}_{{\rm E};\mu} \nonumber \\
  &+&\Big[ \frac{d(\mu_b\delta n )}{d\tau}  
    -(\mu_b \delta n ) [ \frac{1}{\mu_b}\frac{d\mu_b}{d\tau}]  
    -\frac{d\Lambda}{d\tau} \Big]  - \chi \Pi \theta,   \qquad 
\end{eqnarray}
the second law of thermodynamics is guaranteed  
(with $\zeta$, $\lambda$ and $\eta$ being, respectively, 
the bulk viscosity, heat conductivity and shear viscosity - all positive constants), 
\begin{eqnarray}
 TS^{\mu}_{;\mu} = \frac{\Pi^2}{\zeta} -\frac{W^{\mu}W_{\mu}}{\lambda T} 
  +\frac{\pi^{\mu\nu}\pi_{\mu\nu}}{2\eta} \ge 0, 
\end{eqnarray}
provide that the following constitutive equations are hold:   
\begin{subequations}
\begin{eqnarray} 
\frac{\Pi}{\zeta}~ &=& -(1+\chi)\theta 
  -(\frac{\kappa}{\Pi} +2\xi)\frac{d\Pi}{d\tau} 
  +(\frac{\xi}{\mu_b}\frac{d\mu_b}{d\tau}) \Pi , \label{eq:constitutive_Pi}  
\qquad \\
\frac{W_{\mu}}{\lambda T} &=& \frac{\nabla_{\mu}T}{T} -u^{\lambda}u_{\mu;\lambda} -2\xi'\frac{dW_{\mu}}{d\tau}
+(\frac{\xi'}{\mu_b}\frac{d\mu_b}{d\tau}) W_{\mu}, \qquad \\
\frac{\pi_{\mu\nu}}{2\eta} &=& \nabla_{\langle\mu}u_{\nu\rangle} -2\xi''\frac{d\pi_{\mu\nu}}{d\tau} 
+(\frac{\xi''}{\mu_b}\frac{d\mu_b}{d\tau}) \pi_{\mu\nu}. 
\end{eqnarray}
\end{subequations} 
Because eq.(\ref{eq:constitutive_Pi}) includes term proportional to 
$1/\Pi ~d\Pi/d\tau$, one can introduce for the bulk pressure $\Pi$ 
an arbitrary constant $z$ (with dimension [GeV]$^4$) and write: 
\begin{eqnarray}
[1-\gamma_{\Pi} 
 ]\frac{\Pi}{z} 
+2\zeta \xi \frac{d}{d\tau}\frac{\Pi}{z} +\frac{\kappa\zeta}{z}\frac{d}{d\tau}\ln\frac{\Pi}{z}
=-\frac{\zeta}{z} (1+\chi) \theta, \qquad 
\end{eqnarray}
where $\gamma_{\Pi}\equiv \zeta\frac{\xi}{\mu_b}\frac{d\mu_b}{d\tau}$. \\

\subsection{Small perturbations of the dissipative fields}

To check the stability and causality of our model, 
we shall consider small perturbations of $\Pi$, $W^{\mu}$ and $\pi^{\mu\nu}$ fields. 
The bulk pressure $\Pi$ can be written as  $\Pi=\Pi_0+\delta\Pi$ with the background 
reference field $\Pi_0$ and its small perturbation field $\delta\Pi$ field \footnote{
One can also regard $\Pi_0$ as the value of $\Pi$ at initial proper time $\tau_0$. 
Correspondingly, one can write $\theta_0=u^{\mu}_{0;\mu}$, obtained from the initial flow 
vector field $u^{\mu}_{0}$ which is the initial velocity field at $\tau_0$.  
In this case, $\delta \Pi$ can be interpreted as $\delta \Pi = \Pi (\tau) -\Pi (\tau_0)$. 
Therefore, $\Pi_0$ is in fact a kind of parameter showing the degree non-equilibrium at initial stage.}.
Identifying arbitrary constant $z$ with $\Pi_0$ 
and noticing that $\frac{d}{d\tau} \ln (1+\frac{\delta\Pi}{\Pi_0}) 
\approx \frac{1}{\Pi_0}\frac{d}{d\tau}\delta \Pi$, 
one can rewrite the above equation as the following a set of equations:  
\begin{subequations}
\begin{eqnarray}
 && \Pi_0 =  -[\frac{1+\chi}{1-\gamma_{\Pi}}] ~\zeta\theta_0,   \label{eq:background_forPi}\\ 
 && \tau_{\Pi} \frac{d \delta\Pi}{d\tau} +[1-\gamma_{\Pi}]\delta\Pi = -\zeta (1+\chi) \delta\theta \label{eq:linearized_forPi}, 
\end{eqnarray}
\end{subequations}
where the relaxation time $\tau_{\Pi}$ is given by 
\begin{eqnarray}
 \tau_{\Pi} = \zeta \left( 2\xi + \kappa/\Pi_0 \right), 
\end{eqnarray}
and $\theta =\theta_0+\delta\theta$. 
The eq.(\ref{eq:background_forPi}) represents a background $\Pi_0$ field and 
eq.(\ref{eq:linearized_forPi}) represents equation of motion for the perturbed field 
of the bulk pressure $\delta \Pi$ 
\footnote{
It would be interesting to study fluid under the condition of constant bulk pressure 
because there exist several phenomena it can be apply to like, i.e, for example, 
fluid in the huge bag with constant bag pressure (which can be identified with the 
quark-gluon plasma) or matter under cosmological constant 
in the universe, to name a few.}.

Similar approach can be also applied to $W^{\mu}$ and $\pi^{\mu\nu}$,
with perturbation of the heat flow and the shear viscosity around 
$W^{\mu}=0$ and $\pi^{\mu\nu}=0$, leading to 
\begin{eqnarray}
&& \tau_{w} \frac{d \delta W^{\mu}}{d\tau} +[1-\gamma_w]  \delta W^{\mu} 
=\lambda T \Big[ \frac{\nabla^{\mu} T}{T}  - u^{\lambda}u^{\mu}_{;\lambda} \Big],\qquad \\ 
&&\tau_{\pi} \frac{d \delta\pi^{\mu\nu}}{d\tau} +[1-\gamma_{\pi}] \delta\pi^{\mu\nu} =  
2\eta \nabla^{\langle\mu}u^{\nu\rangle}. 
\end{eqnarray}
Here $\tau_w$ and $\tau_{\pi}$ denote the respective relaxation times:  
\begin{eqnarray*} 
  \tau_{w} = 2\xi' \lambda T, \quad 
  \tau_{\pi}= 2\xi'' \eta,  
\end{eqnarray*}
and $\gamma_{w}\equiv \frac{\lambda T\xi'}{\mu_b}\frac{d\mu_b}{d\tau}$, 
$\gamma_{\pi}\equiv \frac{2\eta\xi''}{\mu_b}\frac{d\mu_b}{d\tau}$. 

For simplicity, we drop in what follows terms with 
$\gamma_{\Pi}$, $\gamma_{w}$ and $\gamma_{\pi}$ 
(assuming that $\frac{1}{\mu_b}\frac{d\mu_b}{d\tau} \approx 0$) 
and arrive at the following constitutive equations:  
\begin{subequations}
\begin{eqnarray}
 && \tau_{\Pi} \frac{d \delta\Pi}{d\tau} +\delta\Pi = -\zeta (1+\chi) \delta\theta \label{eq:linearized_forPi2},\\
 && \tau_{w} \frac{d \delta W^{\mu}}{d\tau} +  \delta W^{\mu} 
=\lambda T \Big[ \frac{\nabla^{\mu}T}{T}  - u^{\lambda}u^{\mu}_{;\lambda} \Big],\qquad \\ 
&&\tau_{\pi} \frac{d \delta\pi^{\mu\nu}}{d\tau} + \delta\pi^{\mu\nu} =  
2\eta \nabla^{\langle\mu}u^{\nu\rangle}. 
\end{eqnarray}
\end{subequations}
\section{Stability of the fluid obtained from our model}\label{sec.III} 
The stability of general class of dissipative relativistic fluid theories 
was investigated by Hiscock and Lindblom \cite{PhysRevD.31.725,PhysRevD.35.3723,Hiscock1983466}. 
Denoting by $\delta V(x)$ the difference between the actual non-equilibrium value of a field $V(x)$ and 
the value  in the background reference state, $V_0 (x)$, 
we assume that variations $\delta V$ are small enough so that their evolution is 
adequately described by the linearized equations of motion  describing the background state.  
We shall now investigate the stability of the fluid obtained in our model following prescription 
proposed in Ref. \cite{PhysRevD.31.725,PhysRevD.35.3723,Hiscock1983466}. 
In what follows: 
\begin{enumerate} 
\item 

The background reference state 
is assumed to be homogeneous in space. 
Notice that, unlike in Ref. \cite{PhysRevD.31.725,PhysRevD.35.3723,Hiscock1983466}, 
in our case it is not an equilibrium state but rather a non-equilibrium one with  $\Pi=\Pi_0$ 
and with $W^{\mu}=\pi^{\mu\nu}=0$. 
Furthermore, the background space-time is assumed to be flat Minkowski space, 
so that all background field variables have vanishing gradients. 
\item
We consider following plane wave form of perturbation propagating in $x$ direction 
\begin{eqnarray}
 \delta V = \delta V_0 \exp(ikx +\Gamma \tau). \label{eq:plane-wave} 
\end{eqnarray} 
\end{enumerate}
Linearized equations for dissipative fluid dynamical model are given by   
\begin{subequations}
\begin{eqnarray}
\delta [T^{\mu\nu} ]_{;\mu}&=&0, \\
\delta [N^{\mu}]_{;\mu}&=&0,
\end{eqnarray}
\end{subequations}
with the perturbed energy-momentum tensor and baryon number current given by, respectively: 
\begin{subequations} 
\begin{eqnarray}
 \delta [T^{\mu\nu}] 
 &=& 
(\varepsilon_{\rm eq}^* + P_{\rm eq}^*) (\delta u^{\mu}u^{\nu} + u^{\mu}\delta u^{\nu}) \nonumber \\
 &+& (\delta \varepsilon_{\rm eq}^* +\kappa \delta \Pi) u^{\mu}u^{\nu} 
-(\delta P_{\rm eq}^* +\delta \Pi) \Delta^{\mu\nu} \nonumber \\ 
 &+& \delta W^{\mu} u^{\nu} + \delta W^{\nu} u^{\mu} +\delta \pi^{\mu\nu},\label{eq:linearized_DeltaTmunu} \\ 
 \delta [N^{\mu}]
 &=&  \delta n_{\rm eq}  u^{\mu} + n_{\rm eq}  \delta u^{\mu}. \label{eq:linearized_DeltaN}
\end{eqnarray}
\end{subequations}
Here $\varepsilon_{\rm eq}^*$ and $P_{\rm eq}^{*}$ are energy density and pressure in the background non-equilibrium state 
\begin{eqnarray*}
 \varepsilon_{\rm eq}^* \equiv \varepsilon_{\rm eq} +\kappa\Pi_0 , \quad  
  P_{\rm eq}^{*}\equiv P_{\rm eq}+\Pi_0. \quad 
\end{eqnarray*}
However,  since $\Pi_0={\cal O}(1)$ and $\delta[\Pi_0]=0$ 
(it has vanishing gradient and is constant in $\tau$), terms proportional to 
$\Pi_0$ do not contribute to the linearized equations eq.(\ref{eq:linearized_DeltaTmunu}) 
(with only terms up to the order ${\cal O}(1)$ present).  
Hence, one can replace in the eq.(\ref{eq:linearized_DeltaTmunu}) 
$\varepsilon^*$ and $P^*$ by the, respectively, $\varepsilon_{\rm eq}$ and $P_{\rm eq}$. 

The perturbed fluid dynamical fields must satisfy constraints 
\begin{eqnarray*}
u_{\mu}\delta u^{\mu}=0, \quad \delta W_{\mu} u^{\mu}=0, \quad \delta \pi^{\mu\nu} u_{\mu}=0. 
\end{eqnarray*}
Hence, in the rest frame of fluid, the proper time $\tau$ component of the 
flow velocity field vanishes, $\delta u^{\tau}\equiv 0$. 
Moreover, since the contribution from $\delta n$ is ${\cal O}(2)$, the term $\delta n u^{\mu}$ 
appearing in eq.(\ref{eq:linearized_DeltaN}) 
can be disregarded. We obtain therefore following linearizing equations 
for the energy-momentum tensor and baryon number current:  
\begin{eqnarray}
  \delta [T^{\mu\nu}]_{;\mu} 
  &=& (\varepsilon_{\rm eq} +  P_{\rm eq})((ik\delta u^{x}) u^{\nu} +\Gamma \delta u^{\nu}) \nonumber \\
  &+& (\Gamma \delta \varepsilon_{\rm eq}+\kappa\Gamma\delta \Pi ) u^{\nu} 
    - (\nabla^{\nu} \delta P_{\rm eq}+\nabla^{\nu}\delta \Pi) \nonumber \\
  &+& (ik \delta W^x ) u^{\nu} +\Gamma\delta W^{\nu}+ (ik)\delta\pi^{x\nu} =0, \\
  \delta [N^{\mu}]_{;\mu} &=&  \Gamma \delta n_{\rm eq} +  n_{\rm eq} (ik\delta u^{x}) =0. 
\end{eqnarray}
The linearized constitutive equation for $\Pi$ have now the form (with  $\tilde\kappa\equiv 1+\kappa$ ) : 
\begin{eqnarray}
\frac{(1+\tau_{\Pi}\Gamma )}{\zeta} \delta \Pi 
&=& -\tilde\kappa (ik)\delta u^{x}, \label{eq:Linearized_Pi} 
\end{eqnarray}
whereas for $W^{\mu}$ and $\pi^{\mu\nu}$, we have 
\begin{eqnarray}
&&\frac{(1+\tau_{w}\Gamma )}{\lambda T} \delta W^{\mu} 
  =\frac{\nabla^{\mu} \delta T}{T} -\frac{d\delta u^{\mu}}{d\tau}, \\ 
&& \frac{(1+\tau_{\pi}\Gamma )}{2\eta} \delta \pi^{\mu\nu} =  
   \nabla^{\langle \mu} \delta u^{\nu\rangle} \nonumber \\
&& \qquad 
 =-\frac{1}{2}(ik)[ \delta^{\mu}_{x}\delta u^{\nu} +\delta^{\nu}_x\delta u^{\mu} -\frac{2}{3} \delta^{\mu\nu}\delta u^x]. 
\end{eqnarray}
The parameters $\xi$, $\xi'$ and $\xi''$ introduced in eq.(\ref{eq:extended_matching_2}) 
have been absorbed in the expressions for relaxation time, $\tau_{\Pi}$,$\tau_{w}$ and $\tau_{\pi}$, respectively. 
On the other hand parameter $\kappa$ in the expression of the entropy production is kept not absorbed 
in $\tau_{\Pi}$.
Its role will be discussed later. \\

All perturbation equations can be expressed in concise matrix form: 
\begin{eqnarray}
  M^A_B \delta Y^B=0, \label{eq:Linearized Matrix}
\end{eqnarray}
where $\delta Y^B$ represents the list of fields. 
The system matrix $M^A_B$ can be expressed in a block-diagonal form when one 
chooses the following set of perturbation variables \cite{PhysRevD.31.725} 
\begin{eqnarray}
  \delta Y^B &=&\{~ 
  \delta \varepsilon_{\rm eq}, \delta n_{\rm eq}, \delta u^x, \delta \Pi, \delta W^x,\delta \pi^{xx}, \nonumber \\ 
&& \qquad \delta u^y, \delta W^y,\delta \pi^{xy}, ~\delta u^z, \delta W^z, \delta \pi^{xz}, \nonumber \\
&& \qquad \delta \pi^{yz}, \delta \pi^{yy}-\delta \pi^{zz}  ~\}. 
\end{eqnarray}
In this case, 
\begin{eqnarray}
  {\bf M}= \left( \begin{array}{cccc}
  {\bf Q} & & \\
 & {\bf R} & & \\ 
 & & {\bf R} &  \\
 & & & {\bf I} \\
\end{array}   \right) , 
\end{eqnarray}
where the matrices ${\bf Q}$ and ${\bf R}$ are given by 
\begin{subequations}
\begin{eqnarray}
&&{\bf Q}= \nonumber \\ 
&& \left( 
\begin{array}{cccccc}
          0 & \Gamma & ik n_{\rm eq}  & 0 & 0 & 0 \\ 
    \Gamma & 0 & ikh_{\rm eq} & \kappa\Gamma & ik & 0 \\ 
    ik\frac{\partial P_{\rm eq}}{\partial \varepsilon_{\rm eq}}&
    ik\frac{\partial P_{\rm eq}}{\partial n_{\rm eq}} & \Gamma h_{\rm eq}  
    & ik & \Gamma & ik \\ 
    0 & 0 & ik \tilde\kappa & \frac{1+\tau_{\Pi}\Gamma}{\zeta} & 0 & 0 \\ 
    \frac{ik}{T}\frac{\partial T}{\partial \varepsilon_{\rm eq}}
 & \frac{ik}{T}\frac{\partial T}{\partial n_{\rm eq}} & \Gamma 
 & 0 & \frac{1+\tau_{w}\Gamma}{\lambda T} & 0 \\ 
    0& 0 & ik& 0 & 0 & \frac{1+\tau_{\pi}\Gamma}{4\eta/3} \\ 
\end{array}
\right), \nonumber \\ 
&& \\
&& 
   {\bf R}=\left( 
   \begin{array}{ccc}
    h_{\rm eq} \Gamma & \Gamma & ik \\ 
   \Gamma & \frac{1+\tau_{w}\Gamma}{\lambda T} & 0 \\
    ik & 0 & \frac{1+\tau_{\pi}\Gamma}{\eta} \\
  \end{array} \right), 
\end{eqnarray}
\end{subequations}
respectively, and {\bf I} is the $2\times2$ unit matrix. The $h_{\rm eq}$ denotes the enthalpy density 
which is defined by $h_{\rm eq} \equiv \varepsilon_{\rm eq}+P_{\rm eq}$.\\

For $\Gamma$ and $k$ satisfying dispersion relation 
\begin{eqnarray}
[{\rm det}{\bf M}]=[{\rm det}~{\bf R}]^2[{\rm det}~{\bf Q}]=0
\end{eqnarray} 
one has plane-wave solution such like eq.(\ref{eq:plane-wave}) for the linearized equations of the system 
eq.(\ref{eq:Linearized Matrix}).  

The set of all exponential plane-wave solutions is characterized by the collection of roots obtained by 
setting the determinants of matrix ${\bf R}$ and ${\bf Q}$ separately equal to zero \cite{PhysRevD.31.725,PhysRevD.35.3723}. 

In what follows we shall discuss in detail the stability of transverse and longitudinal modes 
separately and provide the general stability conditions required in our case. 

\subsection{Transverse mode} \label{sec:transverse_mode} 
The dispersion relation obtained by setting  
\begin{eqnarray}
[\eta][\lambda T] 
{\rm det}({\bf R}) \equiv \sum_{m=0}^{3} r_m \Gamma^m =0 
\label{eq:transverse_3rd_equation}
\end{eqnarray}
corresponds to the solution of the perturbation equation which is referred to as 
the so-called transverse mode. The coefficients $r_m$ are given by 
\begin{subequations}
\begin{eqnarray*}
 r_3 &=&  \tau_{w} [h_{\rm eq} \tau_{w} -\lambda T ],  \\ 
 r_2 &=&  h_{\rm eq} (\tau_{w}+\tau_{\pi})-\lambda T,  \\
 r_1 &=&  h_{\rm eq} +\eta\tau_{w}k^2,\\ 
 r_0 &=&  \eta k^2. 
\end{eqnarray*}
\end{subequations} 
When all coefficients of the 3rd order equation (\ref{eq:transverse_3rd_equation}) 
have the same sign, the three solutions consists of a pure real root with negative sign 
and two complex number roots with negative in real parts. 
In this case, since each real part of those solutions is negative, 
the general solution which is linear combination of those three solutions,  
is stable. 
Since $r_0$ and $r_1$ are positive defined, 
then the stability condition sought after is that $r_3$ and $r_2$ must be 
simultaneously positive, i.e., that,  
\begin{subequations}
\begin{eqnarray}
\frac{\lambda T}{\tau_{w}} < h_{\rm eq} \quad\mbox{and}\quad  
\frac{\lambda T}{\tau_{w}+\tau_{\pi}} < h_{\rm eq}. \label{eq:condition_for_transverse} 
\end{eqnarray}
Then, the necessary and sufficient condition which simultaneously satisfies 
eq.(\ref{eq:condition_for_transverse}) is 
\begin{eqnarray}
 \frac{\lambda T}{\tau_{w}} < h_{\rm eq}~. \label{eq:Rmode_stability}
\end{eqnarray} 
\end{subequations}
When condition eq.(\ref{eq:Rmode_stability}) is satisfied, 
the transverse mode solution of the dissipative fluid dynamical equations 
can evolves without facing an instability problem. 

Note that for $\tau_{\Pi}=\tau_{w}=\tau_{\pi}=0$ (with also 
$\kappa$, $\xi$, $\xi'$ and $\xi''$ are all zero but with $\zeta,\lambda,\eta\ne0$, 
i.e., in the case of the standard matching condition), 
one gets the dispersion relation obtained already by Hiscock and Lindblom in Ref.\cite{PhysRevD.31.725}, 
$\Gamma=\frac{h_{\rm eq}}{2\lambda T}\pm \sqrt{(\frac{h_{\rm eq}}{2\lambda T})^2+\frac{\eta k^2}{\lambda T}}$. 
It has two real solutions the linear combination of which leads to well known instability of the transverse mode. 

\subsection{Longitudinal mode}
The determinant of the coefficient matrix ${\bf Q}$ can be expanded in the following way: 
\begin{eqnarray}
{\rm det}({\bf Q}) &=& (ik\Gamma) ~{\rm det}({\bf Q}_1) \nonumber \\ 
&-& [\frac{1+\tau_{\pi}\Gamma}{4\eta/3}] \tilde\kappa(ik)   ~{\rm det}({\bf Q}_2) \nonumber \\
&+& [\frac{1+\tau_{\pi}\Gamma}{4\eta/3}]  [\frac{1+\tau_{\Pi}\Gamma}{\zeta}] ~{\rm det}({\bf Q}_3), \label{eq:det_Q} 
\end{eqnarray}
where 
\begin{subequations}
\begin{eqnarray}
 {\bf Q}_1&=& \left(
\begin{array}{cccc} 
    \Gamma & \kappa\Gamma & ik & 0 \\ 
    ik\frac{\partial P_{\rm eq}}{\partial \varepsilon_{\rm eq}} & ik & \Gamma & ik \\ 
    0 & \frac{1+\tau_{\Pi}\Gamma}{\zeta} & 0 & 0 \\ 
    \frac{ik}{T}\frac{\partial T}{\partial \varepsilon_{\rm eq}} & 0 
 & \frac{1+\tau_{w}\Gamma}{\lambda T} & 0 \\ 
\end{array}
\right), \\
 {\bf Q}_2&=& 
\left(
\begin{array}{cccc}
     0 & \Gamma & 0 & 0 \\ 
    \Gamma & 0 & \kappa\Gamma& ik \\ 
     ik\frac{\partial P_{\rm eq}}{\partial \varepsilon_{\rm eq}}& 
     ik\frac{\partial P_{\rm eq}}{\partial n_{\rm eq}}& ik & \Gamma \\ 
    \frac{ik}{T}\frac{\partial T}{\partial \varepsilon_{\rm eq}}
 & \frac{ik}{T}\frac{\partial T}{\partial n_{\rm eq}} & 0 
 & \frac{1+\tau_{w}\Gamma}{\lambda T} \\  
\end{array} \right), \\
{\bf Q}_3 &=& \left(
\begin{array}{cccc}
     0 & \Gamma & ik n_{\rm eq}   & 0 \\ 
    \Gamma & 0 & ikh_{\rm eq} & ik \\ 
        ik\frac{\partial P_{\rm eq}}{\partial \varepsilon_{\rm eq}}&
        ik\frac{\partial P_{\rm eq}}{\partial n_{\rm eq}} & \Gamma h_{\rm eq}  
    & \Gamma \\ 
        \frac{ik}{T}\frac{\partial T}{\partial \varepsilon_{\rm eq}}
    &  \frac{ik}{T}\frac{\partial T}{\partial n_{\rm eq}} & \Gamma 
    & \frac{1+\tau_{w}\Gamma}{\lambda T} \\  
\end{array} \right). \qquad 
\end{eqnarray}
\end{subequations}
Then the frequencies of the so-called longitudinal mode are given by the roots of the following dispersion relation:  
\begin{eqnarray}
[~\frac{4\eta}{3\tau_{\pi}} \frac{\lambda T}{\tau_{w}} \frac{\zeta}{\tau_{\Pi}} ~] {\rm det}({\bf Q})
  \equiv \sum_{n=0}^{6} q_n \Gamma^n =0 , 
\label{eq:Gamma_polynomial} 
\end{eqnarray}
where the coefficients $q_n$ are given by
\begin{subequations}
\begin{eqnarray*}
 q_6 &=& \frac{\lambda T}{\tau_{w}} -  h_{\rm eq} ,\\
 q_5 &=& \frac{\lambda T}{\tau_{w}} \frac{2}{\tilde\tau_{\Pi\pi}} -\frac{3h_{\rm eq} }{\tilde\tau} ,\\
 q_4 &=& -\Big[~ \frac{\tilde\kappa\zeta}{\tau_{\Pi}}x 
  +\frac{\lambda T}{\tau_{w}} (\kappa \frac{\tilde\kappa\zeta}{\tau_{\Pi}}X -J_2)
  +\frac{4}{3}\frac{\eta}{\tau_{\pi}}  +J_1 \Big]k^2 \nonumber \\
&&+\Big[  \frac{\lambda T -3\bar\tau h_{\rm eq}}{\langle \tau \rangle^3} ~\Big] , \\
 q_3 &=& -\Big[ x\frac{\tilde\kappa\zeta}{\tau_{\Pi}}\frac{2}{\tilde\tau_{w\pi}}
  +\frac{\lambda T}{\tau_{w}} (\kappa  \frac{\tilde\kappa\zeta}{\tau_{\Pi}}\frac{X}{\tau_{\pi}} 
  -\frac{2J_2}{\tilde\tau_{\Pi\pi}}) \nonumber \\
&& +\frac{4\eta}{3\tau_{\pi}}\frac{2}{\tilde\tau_{\Pi w}} +\frac{3J_1}{\tilde\tau} \Big]k^2 -\frac{h_{\rm eq}}{\langle \tau \rangle^3} , \\ 
 q_2 &=& -\frac{\lambda T}{\tau_{w}} \Big[ 
  \frac{\tilde\kappa\zeta}{\tau_{\Pi}}X+\frac{4}{3}\frac{\eta}{\tau_{\pi}}X  - J_3 \Big] k^4 
-\Big[
 \frac{\tilde\kappa\zeta}{\tau_{\Pi}} \frac{x}{\tau_w\tau_{\pi}} \nonumber \\
&&
-\frac{\lambda T}{\tau_w}\frac{J_2}{\tau_{\Pi}\tau_{\pi}}   
+\frac{4\eta}{3\tau_{\pi}} \frac{1}{\tau_{\Pi}\tau_w} 
+\frac{3\bar{\tau}}{\langle\tau\rangle^3}J_1 \Big] k^2,\\ 
 q_1 &=&  -\frac{\lambda T}{\tau_w} \Big[ 
    \frac{\tilde\kappa\zeta}{\tau_{\Pi}} \frac{X}{\tau_{\pi}} 
 +\frac{4\eta}{3\tau_{\pi}} \frac{X}{\tau_{\Pi}}
 -\frac{2J_3}{\tilde\tau_{\Pi\pi}}
  \Big]k^4 -\frac{J_1}{\langle \tau \rangle^3}k^2 ,\qquad \\
 q_0 &=& \Big[ \frac{\lambda T}{\tau_w} \frac{J_3}{\tau_{\Pi}\tau_{\pi}} \Big] k^4.
\end{eqnarray*} 
\end{subequations}
The notations used for $\tilde\tau_{a,b}$(for $a,b=\Pi, w$ and $\pi$), $\tilde\tau$ and $\bar{\tau}$ is:  
\begin{eqnarray*}
 \frac{1}{\tilde\tau_{ab}}&\equiv& \frac{1}{2}(\frac{1}{\tau_a}+\frac{1}{\tau_b}),\quad  
 \frac{1}{\tilde\tau} \equiv \frac{1}{3}(\frac{1}{\tau_{\Pi}}+\frac{1}{\tau_{w}}+\frac{1}{\tau_{\pi}}), \\
 \bar\tau &\equiv& \frac{1}{3}(\tau_{\Pi}+\tau_{w} +\tau_{\pi} ) ,\quad 
 \langle \tau \rangle \equiv (\tau_{\Pi}\tau_{w} \tau_{\pi} )^{1/3}, 
\end{eqnarray*}
whereas $J_n~(n=1,2,3)$ are defined by 
\begin{eqnarray*}
 J_1&\equiv& h_{\rm eq} \frac{\partial P_{\rm eq}}{\partial \varepsilon_{\rm eq}} 
   + n_{\rm eq} \frac{\partial P_{\rm eq}}{\partial n_{\rm eq}}, \\
 J_2&\equiv& \frac{\partial P_{\rm eq}}{\partial \varepsilon_{\rm eq}} 
       + \frac{n_{\rm eq}}{T}\frac{\partial T}{\partial n_{\rm eq}},\\ 
 J_3&\equiv&  \frac{n_{\rm eq}}{T}[
         \frac{\partial P_{\rm eq}}{\partial \varepsilon_{\rm eq}}
         \frac{\partial T}{\partial n_{\rm eq}}
       -\frac{\partial T} {\partial \varepsilon_{\rm eq}}
         \frac{\partial P_{\rm eq}}{\partial n_{\rm eq}}]~. 
\end{eqnarray*}
The definitions of $x$ and $X$ are 
\begin{eqnarray*}
   x\equiv 1-\kappa \frac{\partial P_{\rm eq}}{\partial \varepsilon_{\rm eq}}, \quad 
   X\equiv \frac{1}{T} \frac{\partial T}{\partial \varepsilon_{\rm eq}} . 
\end{eqnarray*}
Notice that $\tau$ (and quantities such as $\tilde\tau$ and $\bar{\tau}$ associated with it) 
are positive. Also $X$ is positive, but the sign of $x$ depends on 
$\frac{\partial P_{\rm eq}}{\partial \varepsilon_{\rm eq}}$ and $\kappa$. 
In what concerns the signs of  $J_1$, $J_2$ and $J_3$ it occurs that, 
apparently, $J_1>0$. On the other hand, because 
$\frac{\partial T}{\partial n_{\rm eq}}\big|_{\varepsilon_{\rm eq}}<0$ 
and other three factors in $J_3$ are positive, then we have $J_3<0$. 
In what concerns $J_2$, since one has  (see Appendix \ref{appendix:B}) 
\begin{eqnarray*}
J_2=\frac{1}{J}[s_{\rm eq} \frac{\partial n_{\rm eq}}{\partial \mu}-n_{\rm eq} \frac{\partial n_{\rm eq}}{\partial T} 
-\frac{n_{\rm eq}}{T} \frac{\partial \varepsilon_{\rm eq}}{\partial \mu}], 
\end{eqnarray*} 
where $J$ is Jacobian $J\equiv \frac{\partial(\varepsilon_{\rm eq}, n_{\rm eq})}{\partial(T,\mu)}>0$, we have $J_2<0$. 
To observe $J_2<0$, consider $J_2$ in the Boltzmann approximation. In the limit, we have  
\begin{eqnarray*}
J_2=\frac{1}{J}\frac{(P_{\rm eq}-\varepsilon_{\rm eq})n_{\rm eq}}{T^2} ,  
\end{eqnarray*}
which has negative sign apparently.

\subsection{General stability conditions} 
To get stable evolution of the fluid, both transverse and longitudinal mode must be simultaneously stable 
against small perturbation and each condition needs to be satisfied in a  without a contradictory. 
The longitudinal mode needs to satisfy the requirement that, as the same as the transverse mode, 
all coefficients $q_n$ in eq.(\ref{eq:Gamma_polynomial}) have the same sign in order to obtain 
solutions of plane-wave type eq.(\ref{eq:plane-wave}) 
with negative real part of complex $\Gamma$, Re~$\Gamma<0$.   
Since $q_0$ (and also $q_1$) is negative then all the coefficients $q_i$ must be negative, $q_i<0$ $(i=2,3,\cdots,6)$. 
Now conditions that $q_6<0, q_5<0$ and the second term of $q_4$ are negative, are equivalent to demanding that 
respectively,  
\begin{eqnarray*} 
\frac{\lambda T}{\tau_w} < h_{\rm eq},\quad 
\frac{\lambda T}{\tau_w} < [1+\frac{1/\tau_w}{1/\tau_{\Pi}+1/\tau_{\pi}}] h_{\rm eq}
\end{eqnarray*}
and
\begin{eqnarray*}  
\frac{\lambda T}{\tau_w} < [1+\frac{\tau_{\Pi}+\tau_{\pi}}{\tau_{w}}] h_{\rm eq}.  
\end{eqnarray*} 
The  necessary and sufficient condition for all these three quantities to be negative is that   
\begin{subequations}
\begin{eqnarray}
  \frac{\lambda T}{\tau_w} < h_{\rm eq}. \label{eq:stability1} 
\end{eqnarray}
This is exactly the same condition as that for the stability of the transverse mode, 
cf. eq.(\ref{eq:Rmode_stability}).  
To continue, conditions that the first term of $q_4$, the $q_3$ and $q_2$ to be all negative 
is that $x>0$, i.e., that 
\begin{eqnarray}
    \frac{\partial P_{\rm eq}}{\partial \varepsilon_{\rm eq}}\bigg|_{n_{\rm eq}} 
    = c_s^2 +\frac{1}{T}\frac{\partial P_{\rm eq}}{\partial s_{\rm eq}}  \bigg|_{\varepsilon_{\rm eq}} \le~ \frac{1}{\kappa}, \label{eq:stability2} 
\end{eqnarray}
\end{subequations}
where $c_s^2\equiv \frac{\partial P_{\rm eq}}{\partial \varepsilon_{\rm eq}}\big|_{s_{\rm eq}}$ is 
adiabatic velocity of sound.  
From eq.~(\ref{eq:stability2}) it is clear that parameter $\kappa$ is related to the speed of sound wave, $c_s$, propagating 
in the fluid. 
Note that, when $\kappa\to 0$ in eq.(\ref{eq:stability2}), one finds that bound on  
the speed of sound, $c_s < 1$,  can be exceeded violating causality. 
On the other hand $c_s \ge 0$. Therefore  
eq.(\ref{eq:stability2}) gives the following restriction on the parameter $\kappa$:  
\begin{eqnarray}
    \frac{1}{T}\frac{\partial P_{\rm eq}}{\partial \varepsilon_{\rm eq}} \le \frac{1}{\kappa} \le 
1+ \frac{1}{T}\frac{\partial P_{\rm eq}}{\partial \varepsilon_{\rm eq}}~. 
\end{eqnarray}

To summarize this part, the stability condition for a dissipative fluid in an 
off-equilibrium state with pressure $P=P_{\rm eq}+\Pi_0$ (the corresponding 
energy density is given by $\varepsilon=\varepsilon_{\rm eq}+\kappa \Pi_0$) 
is that 
\begin{eqnarray}
 \frac{\lambda T}{\tau_w} \le \varepsilon_{\rm eq} + P_{\rm eq} \quad \mbox{and} \quad 
 c_s^2 \le \frac{1}{\kappa} - \frac{1}{T}\frac{\partial P_{\rm eq}}{\partial s_{\rm eq}}\bigg|_{\varepsilon_{\rm eq}}. \quad  
 \label{eq:final_condition0} 
\end{eqnarray} 
This means that the small thermal conduction, 
$\lambda$ and/or large relaxation time, $\tau_w$, is required 
in order to stabilize relativistic dissipative fluid. 
Recalling definition of the relaxation time for the heat flow, $\tau_w\equiv 2\xi'\lambda T$, 
one of the stability conditions, eq.(\ref{eq:stability1}),  
can be expressed as a constraint on the $\xi'$ appearing in eq.(\ref{eq:extended_matching_2}): 
\begin{eqnarray*}
1/\xi'\le 2(\varepsilon_{\rm eq}+P_{\rm eq}).
\end{eqnarray*}  
Thus, conditions eq.(\ref{eq:final_condition0}) for the stability of dissipative fluid can be expressed as the constraints for the 
infinitesimal quantities ${\cal O}(1)$ and ${\cal O}(2)$  
(See eq.(\ref{eq:extended_matching_1}) and (\ref{eq:extended_matching_2})) in the following form:
\begin{subequations} 
\begin{eqnarray}
  && \frac{1}{2\xi'} \le \varepsilon_{\rm eq} + P_{\rm eq}, \label{eq:final_condition1}\\ 
  && \frac{1}{T}\frac{\partial P_{\rm eq}}{\partial s_{\rm eq}} \le \frac{1}{\kappa} \le 1+ \frac{1}{T}\frac{\partial P_{\rm eq}}{\partial s_{\rm eq}}. 
  \label{eq:final_condition2} 
\end{eqnarray}
\end{subequations} 

\section{Summary and concluding remarks}\label{sec.IV}
We have proposed a new formulation of the relativistic dissipative hydrodynamical model 
in the Eckart frame by relaxing the standard matching conditions, 
\begin{eqnarray*}
u_{\mu} u_{\nu} \delta T^{\mu\nu} =0 \quad\mbox{and}\quad u_{\mu}\delta N^{\mu}=0, 
\end{eqnarray*}
and replacing by a more general form, 
\begin{eqnarray*}
 u_{\mu}u_{\nu} \delta T^{\mu\nu} =\Lambda \quad\mbox{and}\quad u_{\mu}\delta N^{\mu}=\delta n. 
\end{eqnarray*} 
We assume that $\Lambda ={\cal O}(1)$, $\delta n ={\cal O}(2)$ and that 
they are given by a simple function of dissipative quantities $\Pi$, $W^{\mu}$ and $\pi^{\mu\nu}$, 
cf. eq.(\ref{eq:extended_matching}). \\

Introducing the extended matching condition, we have also accordingly generalized 
the form of the entropy current for a non-equilibrium state in order that 
the extended matching conditions might be suited (cf. eq.(\ref{eq:entropy_current_our_model})):
\begin{eqnarray*}
   S^{\mu}= S^{\mu}_{\rm eq} +W^{\mu}/T -\alpha \delta n u^{\mu} +\Lambda \beta^{\mu} -\chi\beta\Pi u^{\mu}. 
\end{eqnarray*}
The phenomenological parameter $\chi$ introduced in the generalization of the entropy current form can be fixed 
by the extended thermodynamical stability condition (cf. eq.(\ref{eq:StabilityCondition}))
\begin{eqnarray*}
   -\mu_b\delta n +\Lambda - \chi \Pi =0 . 
\end{eqnarray*}

As the result, the new the entropy current is formally the same as that obtained from Eckart theory (see eq.(\ref{eq:entropy_current})). 
However, our matching conditions change the entropy production given by eq.(\ref{eq:entropy_production}) 
and therefore they also change the equation of motion of fluid 
(i.e., the constitutive equations  $\Pi$, $W^{\mu}$ and $\pi^{\mu\nu}$). 
This is because now $\chi$ appears in the expression for the entropy production 
(cf. the 1st line of eq.(\ref{eq:entropy_production}) 
or, alternatively, in $\Lambda$ and $\delta n$ 
cf. the 2nd line of eq.(\ref{eq:entropy_production})). 
It means then that, because of thermodynamic stability condition, the quantities 
of ${\cal O}(1)$ and ${\cal O}(2)$ 
introduced by the extended conditions change both the entropy production 
and the equation of fluid motion. \\
 
We check the stability of the fluid obtained by our model. 
Small perturbation imposed to a non-equilibrium state characterized by  
a constant bulk pressure $\Pi_0$ and $W^{\mu}=0, \pi^{\mu\nu}=0$ 
as the background reference fields. 
We found that the evolution of fluid is stable against small perturbations 
and the velocity of sound satisfies the usual limits, $0\le c_s \le 1$, 
once eq.(\ref{eq:final_condition1}) and eq.(\ref{eq:final_condition2}) are satisfied. 
Those conditions give restrictions to the entropy production, eq.(\ref{eq:entropy_production}), 
via the constraints for $\kappa$ and $\xi'$ in the extended matching conditions  
eq.(\ref{eq:extended_matching_1}) and (\ref{eq:extended_matching_2}). 

We conclude then, that when the matching condition 
$u_{\mu}u_{\nu}\delta T^{\mu\nu}={\cal O}(1)$ and $u_{\mu}\delta N^{\mu}={\cal O}(2)$ 
are imposed properly,  
the relativistic dissipative fluid described by the Eckart's first order theory 
evolves in stable and causal way.

\acknowledgments
The author would like to thank Grzegorz Wilk for critical reading of this manuscript.

\appendix
\section{Determinant of the matrix ${\bf Q}$}\label{appendix:det_Q}
The first term of eq.(\ref{eq:det_Q}) 
times $ [\frac{4\eta}{3\tau_{\pi}} \frac{\lambda T}{\tau_{w}} \frac{\zeta}{\tau_{\Pi}} ] $ is given by 
\begin{eqnarray}
&& (ik\Gamma) {\rm det}({\bf Q}_1) 
\times [\frac{4\eta}{3\tau_{\pi}} \frac{\lambda T}{\tau_{w}} \frac{\zeta}{\tau_{\Pi}} ] \nonumber \\ 
&&\quad= -\frac{4\eta}{3\tau_{\pi}} k^2\Gamma^4 
      -\frac{4\eta}{3\tau_{\pi}} \frac{2}{\tilde \tau_{\Pi w}} k^2\Gamma^3 \nonumber \\
&& \quad 
      -\frac{4\eta}{3\tau_{\pi}} [\frac{1}{\tau_{\Pi}\tau_{w}}+\frac{\lambda T}{\tau_w} X k^2 ] k^2 \Gamma^2 
      -\frac{4\eta}{3\tau_{\pi}} \frac{\lambda T}{\tau_{w}} \frac{X}{\tau_{\Pi}} k^4\Gamma,\quad \label{eq:appendix1}
\end{eqnarray}
where $X\equiv \frac{1}{T}\frac{\partial T}{\partial \varepsilon_{\rm eq}}$.
The second term is 
\begin{eqnarray}
&& -[\frac{1+\tau_{\pi}\Gamma}{4\eta/3}] 
    \tilde\kappa(ik) ~{\rm det}({\bf Q}_2) 
\times [\frac{4\eta}{3\tau_{\pi}} \frac{\lambda T}{\tau_{w}} \frac{\zeta}{\tau_{\Pi}} ] \nonumber \\
&&\quad = -\tilde\kappa\frac{\zeta}{\tau_{\Pi}} [ \kappa\frac{\lambda T}{\tau_{w}}  X  +x ]k^2\Gamma^4  
- \tilde\kappa\frac{\zeta}{\tau_{\Pi}}  [ \kappa\frac{\lambda T}{\tau_{w}}  \frac{X}{\tau_{\pi}}   
     +\frac{2x}{\tilde \tau_{w \pi}}]k^2 \Gamma^3 \nonumber \\
&& \quad - \tilde\kappa\frac{\zeta}{\tau_{\Pi}} [ \frac{\lambda T}{\tau_{w}} X k^2 + \frac{x}{\tau_{w}\tau_{\pi}} ] k^2\Gamma^2 
- \tilde\kappa \frac{\zeta}{\tau_{\Pi}} \frac{\lambda T}{\tau_{w}} \frac{X}{\tau_{\pi}}  k^4  \Gamma, \label{eq:appendix2}
\end{eqnarray}
where $x\equiv 1-\kappa\frac{\partial P_{\rm eq}}{\partial \varepsilon_{\rm eq}}$. 
The third term is given by 
\begin{eqnarray}
&& [\frac{1+\tau_{\pi}\Gamma}{4\eta/3}]  [\frac{1+\tau_{\Pi}\Gamma}{\zeta}] 
{\rm det}({\bf Q}_3) \times [\frac{4\eta}{3\tau_{\pi}}\frac{\lambda T}{\tau_{w}} \frac{\zeta}{\tau_{\Pi}}] \nonumber \\ 
&&\quad= 
[\frac{\lambda T}{\tau_w} - h_{\rm eq}] ~\Gamma^6 
+ [\frac{\lambda T}{\tau_{w}} \frac{2}{\tilde \tau_{\Pi\pi}}
   -\frac{3}{\tilde \tau} h_{\rm eq}] ~\Gamma^5 
\nonumber \\&&\quad 
+ [\frac{\lambda T}{\tau_w}(\frac{1}{\tau_{\Pi}\tau_{\pi}}+ J_2 k^2) 
  -  J_1 k^2 -\frac{3\bar{\tau}}{\langle \tau \rangle^3} h_{\rm eq} ] ~\Gamma^4  \nonumber \\
&&\quad + [\frac{\lambda T}{\tau_w} \frac{2}{\tilde \tau_{\Pi\pi}}  J_2 k^2  
 -\frac{3}{\tilde\tau} J_1 k^2 -\frac{h_{\rm eq}}{\langle \tau \rangle^3}   ] ~\Gamma^3 \nonumber \\
&&\quad+[\frac{\lambda T}{\tau_w} ( J_3 k^4+ \frac{J_2}{\tau_{\Pi}\tau_{\pi}}k^2  )  
  -\frac{3\bar{\tau}}{\langle\tau\rangle^3} J_1 k^2] ~\Gamma^2 \nonumber \\
&&\quad +[\frac{\lambda T}{\tau_w} \frac{2J_3}{\tilde \tau_{\Pi\pi}}  k^4
             - \frac{J_1}{\langle \tau \rangle^3}k^2] ~\Gamma 
+ \frac{\lambda T}{\tau_w} \frac{J_3}{\tau_{\Pi}\tau_{\pi}} k^4. \label{eq:appendix3} 
\end{eqnarray}
By adding eq.(\ref{eq:appendix1}), eq.(\ref{eq:appendix2}), and eq.(\ref{eq:appendix3}), 
one obtains expression of eq.(\ref{eq:Gamma_polynomial}).  

\section{Sign of $J_2$}\label{appendix:B} 
In order to observe the sign of 
\begin{eqnarray*}
   J_2 =\frac{\partial P_{\rm eq}}{\partial \varepsilon_{\rm eq}} \bigg|_{n_{\rm eq}} 
        -\frac{n_{\rm eq}}{T}\frac{\partial T}{\partial n_{\rm eq}} \bigg|_{\varepsilon_{\rm eq}}, 
\end{eqnarray*}  
consider the following thermodynamical relations 
\begin{eqnarray}
   dP_{\rm eq} = s_{\rm eq} dT + n_{\rm eq} d\mu . \label{eq:appenxi_B1}
\end{eqnarray}
Then, let's change of the thermodynamical variables from $(\varepsilon_{\rm eq}, n_{\rm eq})$ to $(T,\mu)$; 
\begin{subequations}
\begin{eqnarray}
   d\varepsilon_{\rm eq} &=& \frac{\partial \varepsilon_{\rm eq}}{\partial T} dT +  \frac{\partial \varepsilon_{\rm eq}}{\partial \mu} d\mu, 
   \label{eq:appendixB_01}\\
   d n_{\rm eq}  &=& \frac{\partial n_{\rm eq}}{\partial T}  dT + \frac{\partial n_{\rm eq}}{\partial \mu }  d\mu.  
   \label{eq:appendixB_02}
\end{eqnarray}
\end{subequations}
With using a Jacobian defined by  
\begin{eqnarray}
 J\equiv \frac{\partial \varepsilon_{\rm eq}}{\partial T} \frac{\partial n_{\rm eq}}{\partial \mu } -
 \frac{\partial \varepsilon_{\rm eq}}{\partial \mu}  \frac{\partial n_{\rm eq}}{\partial T} , 
\end{eqnarray}
one can inversely express relations eq.(\ref{eq:appendixB_01}) and (\ref{eq:appendixB_02})  as 
\begin{subequations}
\begin{eqnarray}
  dT &=&  \frac{1}{J} \left[ ~\frac{\partial n_{\rm eq}}{\partial \mu }d\varepsilon_{\rm eq} 
  ~- \frac{\partial \varepsilon_{\rm eq}}{\partial \mu} d n_{\rm eq} \right], \\
  d\mu &=& \frac{1}{J}\left[ -\frac{\partial n_{\rm eq}}{\partial T} d\varepsilon_{\rm eq} 
  +\frac{\partial \varepsilon_{\rm eq}}{\partial T} d n_{\rm eq} \right] .
\end{eqnarray}
\end{subequations} 
Substituting the above relations into eq.(\ref{eq:appenxi_B1}), we have
\begin{eqnarray}
  dP_{\rm eq} &=& \left( 
    \frac{s_{\rm eq}}{J} \frac{\partial n_{\rm eq}}{\partial \mu} 
  -\frac{n_{\rm eq}}{J} \frac{\partial n_{\rm eq}}{\partial T} 
  \right) d\varepsilon_{\rm eq}\nonumber \\ 
  &+& 
  \left( 
  -\frac{s_{\rm eq}}{J} \frac{\partial \varepsilon_{\rm eq}}{\partial \mu} 
  +\frac{n_{\rm eq}}{J} \frac{\partial \varepsilon_{\rm eq}}{\partial T}
  \right) dn_{\rm eq} . 
\end{eqnarray} 
Then we have 
\begin{eqnarray}
  \frac{\partial P_{\rm eq}}{\partial \varepsilon_{\rm eq}} \bigg|_{n_{\rm eq}} = 
  \frac{s_{\rm eq}}{J} \frac{\partial n_{\rm eq}}{\partial \mu} 
  -\frac{n_{\rm eq}}{J} \frac{\partial n_{\rm eq}}{\partial T} . 
  \label{eq:appendixB_final1} 
\end{eqnarray}
Eliminating $\mu$ from eq.(\ref{eq:appendixB_01}) and eq.(\ref{eq:appendixB_02}),  we have 
\begin{eqnarray}
  \frac{\partial n_{\rm eq}}{\partial \mu } d\varepsilon_{\rm eq} - \frac{\partial \varepsilon_{\rm eq}}{\partial \mu } dn_{\rm eq} =J dT, 
\end{eqnarray}
and from this, one obtains 
\begin{eqnarray}
  \frac{\partial T}{\partial n_{\rm eq}}\bigg|_{\varepsilon_{\rm eq}} = -\frac{1}{J}\frac{\partial \varepsilon_{\rm eq}}{\partial \mu} . 
  \label{eq:appendixB_final2} 
\end{eqnarray}
Using eq.(\ref{eq:appendixB_final1}) and (\ref{eq:appendixB_final2}), we have an expression 
\begin{eqnarray*}
  J_2=\frac{1}{J}[s_{\rm eq} \frac{\partial n_{\rm eq}}{\partial \mu}-n_{\rm eq} \frac{\partial n_{\rm eq}}{\partial T} 
-\frac{n_{\rm eq}}{T} \frac{\partial \varepsilon_{\rm eq}}{\partial \mu}] .  
\end{eqnarray*}
\bibliography{beyondIS111105}
\end{document}